# The argument for an objective wave function collapse: Why spontaneous localization collapse or no-collapse decoherence cannot solve the measurement problem in a subjective fashion

Fred H. Thaheld
fthaheld@directcon.net

**Abstract**

A more detailed analysis of the measurement problem continues to support the position taken by Shimony and the author that collapse of the wave function takes place in an objective manner in the rhodopsin molecule of the retina.  This casts further doubts on the theories involving a spontaneous localization collapse process or a no-collapse decoherence process taking place in the visual cortex in a subjective fashion.  The possibility is then raised, as per Anandan, as to whether the solution of the measurement problem in quantum theory allows one to address the problem of quantizing gravitation.

**Introduction**

This analysis is a continuation, in a more detailed fashion, of a previous paper addressing the question of the measurement problem (1), and also of specific critiques of the subjective spontaneous localization collapse theory of Ghirardi (2-4), and the subjective no-collapse decoherence theory of Schlosshauer (5,6), both supposedly taking place somewhere in the visual cortex of the brain.  Before commencing this analysis I would like to make clear that in the spirit of Bell (7), I do not deny that there are more or less 'measurement-like' processes going on more or less all the time, more or less everywhere without human intervention.  It is the measurement problem as it has been debated for over 7 decades, involving human subjects or observers, that is the subject of this paper.  My purpose then is to show, in a clearer fashion, why wave function collapse takes place in the rhodopsin molecule as proposed by Shimony (8) and the author (1).

**Analysis**

Each mammalian retina (~250 μm thick) contains $10^8$ photoreceptor rods, and each rod contains ~$10^8$ rhodopsin molecules of a mesoscopic size (9). Rhodopsin, the visual pigment in rod cells, is the covalent complex of a large protein opsin, and a small light-absorbing compound retinal (10). The absorption of light by retinal causes a change in the three-dimensional structure of rhodopsin. The rods are 100 μm in length x 10 μm in dia, with an active outer segment 50 μm long, containing the $10^8$ rhodopsin molecules. The rest of the 50 μm length of the rod is taken up by the inner segment which contains the cell's nucleus and most of its biosynthetic machinery, and by the synaptic terminal which makes contact with the photoreceptor's target cells (10).

One photon can activate only one rhodopsin molecule / rod each time. Thus, the two branches of the wave functions of a superposed photon state cannot simultaneously activate a single rhodopsin molecule in a single rod, nor can they simultaneously activate two rhodopsin molecules in a single rod. Each rod has a minimum quantum detection efficiency of 25% and a maximum quantum detection efficiency of 36%. I.e., *only* 25%-36% of photons incident on the rods are *absorbed*. In addition, these *absorbed* photons produce detectable electrical output signals with a quantum efficiency of 65% for isomerization of retinal in the rhodopsin molecule (9,10).

To help put this in perspective, ~50% of all the photons initially incident upon the cornea are lost through environmental decoherence, up to the point where the retina begins, with ~80% of the remaining photons being lost within the retina.

When one is dealing with non-superposed photon states, the probability p that a rod will absorb a photon is $.25 \leq p \leq .36$. However, when one is dealing with a superposed photon state with two branches or wave functions, and one assumes simultaneity (or close to it in fs) of absorption of these two branches of the wave function by two rods (since the rods work independently and neither is influenced by the other), the probability of success is $p^2$, which translates to $.0625 \leq p^2 \leq .1296$. You can easily see the dramatic loss of superposed states even at this early stage (11), with many more barriers yet to come (10).

We shall now explore the next phase in this process, resulting in further losses of the remaining superposed photon states, which is the issue of the 65% quantum efficiency involved in the photoisomerization of retinal in the rhodopsin molecule. Rhodopsin molecules are so resistant to decoherence (contrary to Schlosshauer's contention (2)), that there are only two ways that they can be activated. First, by successful impingement of a photon. Second, with excitation originating from thermal isomerization of rhodopsin. These are occasional and spontaneous discrete events that resemble single photon responses, that occur once every 90 seconds in a mammalian rod. Since each mammalian rod contains $\sim 10^8$ rhodopsin molecules, each rhodopsin molecule activates spontaneously only *once* every 300 years! This corresponds to an energy barrier of 22 kcal / mol, which is essential for a low noise level. The optimum wavelength of the photons is 500 nm, which is 57 kcal / mol, which represents a 40% efficient utilization of available energy in setting the signal-to-noise ratio (12,13).

It could well be that this energy barrier of the rhodopsin molecule of 22 kcal / mol is where the wave function collapse (nature's measurement or transfer of information) takes place, in the *same place every time for every photon*, in the retinal of the rhodopsin molecule, thereby explaining the consistency of results for all observers of similar events, measurement after measurement. It is important at this juncture to stress that the microscopic or quantum mechanical photon is not impinging upon a *macroscopic* detector or apparatus (which is the usual case) but, upon a more *contiguous mesoscopic* detector, namely the retinal of rhodopsin, leading in a sequential fashion to a macroscopic amplification of the nervous signals. This whole process seems much more *natural*, *seamless* and *less abrupt* than how we usually visualize the collapse of the wave function upon measurement. And, even though the author is an advocate of a wave function collapse mechanism, since the mesoscopic retinal detector is in a *poised* energy state, just waiting to be *triggered* by any photon possessing an appropriate wavelength, might this not also allow the adherents of a *fast* decoherence mechanism some 'wiggle room'?

This would mean that the Heisenberg 'cut' has to follow a very close distance behind, also in the same place every time, either somewhere within the several conformational changes which take place within the rhodopsin molecule or, as a result of the amplification process of the nervous signals (1,6).

One of the major problems in trying to advocate the maintenance of a superposed state, is that the remaining superposed states and their wave functions, after having been subjected to the stochastic decoherence processes outlined above, will arrive at the rods with each photon's wave functions in a multitude of different phases. It would be impossible for all these different superposed states to continue to be maintained after absorption by the rods and being transduced and amplified into the form of detectable electrical output signals (2,4). The same reasoning also applies to the maintenance of a superposed *cis-trans* state at a 40 kDa molecular mesoscopic level, which is ruled out by the above sequential process in an even more stringent fashion (5,6,10). Perusal of Fig.

26-6 from Ref. 10, best illustrates why neither the spontaneous localization collapse theory (2) or the minimal no-collapse decoherence theory (5) are tenable.

**Anandan addendum**

After having attempted to address one of the most difficult and divisive foundational issues in quantum mechanics, this would have normally been the point to end this paper. But, I happened to glance up and saw once again on the corner of my desk, a very nice paper by Jeeva Anandan (which I have read many times), with the fascinating title "Quantum measurement problem and the possible role of the gravitational field" (14). And, recalling Feynman's comment, "…we do not know where we are stupid until we stick our necks out", decided to continue a little further and see what might develop.

Anandan stated that two of the most important unsolved problems in theoretical physics are the problem of quantizing gravitation and the measurement problem in quantum theory and, that it is possible that the solution of each one needs the other. That protective observation suggests that the wave function is real, and therefore the reduction of the wave packet during measurement is a real objective process. He then mentions Penrose as advocating the use of the gravitational field of the wave function to explain its reduction during measurement (15). Penrose's idea is that there may be a nonlinear dynamical mechanism that forbids quantum superpositions of ("too different") space-times.

The microscopic system being observed has its states in the Hilbert space, whose geometry is different than the geometry for space-time into which the gravitational field is incorporated. The wave reduction, or collapse process, brings these two different geometries into contact with each other because of the formation of events when the Hilbert state vector is observed (16,17), suggesting that the gravitational field may be involved in this process. If the gravitational field, which is intimately connected with space-time, causes the reduction of the wave packet, this may explain why the states into which the collapse takes place have a well defined space-time description. This argument suggests that it is not necessary to go down to the scale of Planck length for quantum gravitational effects to become important, because the problem of relating Hilbert space geometry to space-time geometry, which is required by the reduction of the wave packet, exists even at much bigger length scales (14).

**Discussion and conclusion**

1. It is shown, based upon an analysis of detailed research on the retina over a period of several decades, that the theories proposing a possible subjective resolution of the measurement problem by Ghirardi (2) or Schlosshauer (5) are flawed. The basic problem is that *no* superposed photon states can be converted and maintained beyond the retina, either as superposed nervous signals as per Ghirardi, or as superposed *cis-trans* states of the rhodopsin molecule as per Schlosshauer. That the sheer complexity of the retina automatically rules this out.

2. This research lends support to the theory advanced by Shimony and the author, that the superposed photon states most likely collapse in a much simpler objective fashion in the mesoscopic rhodopsin molecule, either as a natural result of the conformational change of the retinal from *cis-trans* or following a process of amplification of the resulting transduced electrical signal.

3. That the exact location of this collapse can be pinpointed within an area ~50 μm in length of the active outer segment of the photoreceptor rods, which are ~100 μm long.

4. This then means, by virtue of pinpointing the area for collapse, that the Heisenberg 'cut' will probably follow a close distance behind, either in the latter portion of the active outer segment or, in the remaining inner segment of 50 μm in length of the photoreceptor rod.

5. This theory has the further advantage that it can be subjected to experiment (4) in the very near future, which is not the case with the competing theories.

6. This will solve two problems at once. The first regarding whether the measurement of a visual field happens in the retina or further up in the visual cortex. And second, whether the reduction mechanism is governed by nonlinear evolution laws.

7. This raises a further tantalizing possibility of exploring Penrose's theory regarding the use of the gravitational field of the wave function to explain its reduction during measurement.

**References**

(1) F.H. Thaheld, 2005. Does consciousness really collapse the wave function?: A possible objective biophysical resolution of the measurement problem. BioSystems 81, 113-124. quant-ph/0509042.
(2) G. Ghirardi, 1999. Quantum superpositions and definite perceptions: envisaging new feasible experimental tests. Phys. Lett. A 262, 1-14.
(3) F. Thaheld, 2000. Comment on "Quantum superpositions and definite perceptions: envisaging new experimental tests". Phys. Lett. A 273, 232-234.
(4) F.H. Thaheld, 2003. Can we determine if the linear nature of quantum mechanics is violated by the perceptual process? BioSystems 71, 305-309.
(5) M. Schlosshauer, 2006. Experimental motivation and empirical consistency in minimal no-collapse quantum mechanics. Ann. Phys. 321, 112-149. quant-ph/0506199.

(6) F.H. Thaheld, 2006. Comment on "Experimental motivation and empirical consistency in minimal no-collapse quantum mechanics". quant-ph/0602190.
(7) J.S. Bell, 1990. Against measurement. Phys. World August, 33-40.
(8) A. Shimony, 1998. Comments on Leggett's "Macroscopic Realism", in *Quantum Measurement: Beyond Paradox.* R.A. Healey and G. Hellman, eds. Univ. of Minnesota, Minneapolis, p. 29.
(9) F. Rieke, D.A. Baylor, 1998. Single-photon detection by rod cells of the retina. Rev. Modern Phys. 70, 1027-1036.
(10) E.R. Kandel, J.H. Schwartz, T.M. Jessell, 2000. *Principles of Neural Science.* $4^{th}$ edn. McGraw-Hill, New York. pp. 507-522. (See especially p. 511, Fig. 26-3 and p. 515, Fig. 26-6.
(11) J. Graham, 2006. Personal communication.
(12) D.A. Baylor, B.J. Nunn, J.L. Schnapf, 1984. The photocurrent, noise and spectral sensitivity of the monkey *Macaca fasciularis.* J. Physiol. (London) 357, 576-607.
(13) J. Nathans, 1992. Rhodopsin: Structure, function and genetics. Biochemistry 31, 4921-4931.
(14) J. Anandan, 1998. Quantum measurement problem and the possible role of the gravitational field. gr-qc/9808033.
(15) R. Penrose, 1996. On gravity's role in quantum state reduction. Gen. Rel. and Grav. 28, 581-600.
(16) J. Anandan, 1980. On the hypotheses underlying physical geometry. Foundations of Physics 10, 601-629.
(17) J. Anandan, 1991. A geometric approach to quantum mechanics. Foundations of Physics 21, 1265-1284.